\documentclass{aa}
\usepackage{graphicx}

\begin{document}

\title{Mass loss rate determination of southern OB stars}
\author{P. Benaglia\inst{1,}\thanks{Member of  Carrera del Investigador,
CONICET},
  C.E. Cappa\inst{1,2,\star}, and
  B.S. Koribalski\inst{3} }

\offprints{P. Benaglia}

\institute{Instituto Argentino de Radioastronom\'{\i}a,
   C.C.5, (1894) Villa Elisa, Buenos Aires, Argentina
   \and
   Fac. de Cs. Astron\'omicas y Geof\'{\i}sicas, La Plata National
   University, Paseo del Bosque S/N, (1900) La Plata, Argentina
   \and
   Australia Telescope National Facility, CSIRO,
   P.O. Box 76, Epping, NSW 1710, Australia}

\date{Received  ; accepted }

\abstract{
 A sample of OB stars (eleven Of, one O and one B
supergiant) has been surveyed with the Australia Telescope Compact
Array at 4.8 and 8.64 GHz with a resolution of
$\sim$2\arcsec--4\arcsec. Five stars were detected; three of them
have negative spectral indices, consistent with non-thermal
emission, and two have positive indices. The thermal radiation
from HD 150135 and HD 163181 can be explained as coming from an
optically thick ionized stellar wind. The non-thermal radiation
from CD--47$^{\circ}$ 4551, HD 124314 and HD 150136 possibly comes
from strong shocks in the wind itself and/or in the wind colliding
region if the stars have a massive early-type companion. The
percentage of non-thermal emitters among detected O stars has
increased up to $\sim$50\%. The Of star HD 124314 clearly shows
flux density variations. Mass loss rates (or upper limits) were
derived for all the observed stars and the results compared with
non-radio measurements and theoretical predictions.
}

\titlerunning{Mass loss rate of OB stars}
\authorrunning{P. Benaglia et al.}
\maketitle

\markboth{P. Benaglia et al.: Mass loss rate of OB stars}{}

\section{Introduction}

Of stars are characterized by N[{\sc iii}] $\lambda\lambda$
4630--34\,\AA\ emission lines and He[{\sc ii}] $\lambda$
4686\,\AA\ emission or absorption lines in their spectra.
According to Maeder (1990) a massive O star evolves into an Of
star which, in turn, goes through the blue supergiant and luminous
blue variable phases, and after that becomes a Wolf Rayet (WR)
star. In addition, Of stars have very strong winds with gas
terminal velocities up to 3000 km s$^{-1}$ (Chlebowski \& Garmany
1991). The velocities, derived from P Cygni profiles, are
appreciably larger than the stellar escape velocity, proving that
important mass loss is taking place in Of stars. Typical values
for the mass loss rate, $\dot{M}$, of Of stars are $\la 10^{-5}$
M$_{\odot}$\,yr$^{-1}$ (Lamers et al. 1995).

Like for WR stars, the action of the wind on the ambient
interstellar medium produces a cavity of hot rarefied gas
surrounded by a slowly expanding envelope. The shells can be
detected at optical and radio wavelengths (Lozinskaya 1982, Cappa
\& Benaglia 1998, Benaglia \& Cappa 1999). Using high resolution,
the wind region itself is observable in the radio continuum
emission. The stellar mass loss rate (or an upper limit) can be
estimated from the radio flux density. Accurate determinations of
$\dot{M}$ are essential in modeling the stellar evolution,
describing the Of-stellar wind and changes in the ambient gas
caused by the mass outflow. Since radio-determined mass loss rates
turned out to be among the most reliable ones, they are used to
calibrate $\dot{M}$ obtained with other methods, i.e. involving
H$\alpha$ equivalent widths or UV profiles.

It is commonly assumed that the dominant driving mechanism for the
winds is radiation pressure. However, $\dot{M}$  values predicted
by radiation-driven wind models differ up to a factor of two with
the observed mass loss rates for O-type supergiants (Vink et al.
2000). Puls et al. (1996) found the discrepancies between observed
and predicted mass loss rates diminished when considering more
accurate input parameters of the involved atomic physics (see
Kudritzki \& Puls 2000 for a comprehensive review). In this
respect, new results on $\dot{M}$ are crucial for testing the
existing wind models and theoretical predictions.

Using the Very Large Array (VLA), systematic observations of O, B
and WR stars were carried out by Abbott et al. (1986) and Bieging
et al. (1989), while more recently Scuderi et al. (1998) also
observed Of as well as O and B-type stars. With the Australia
Telescope Compact Array (ATCA\footnote{The
    Australia Telescope is funded by the Commonwealth of Australia for
    operation as a National Facility managed by CSIRO.}),
Leitherer, Chapman \& Koribalski (1997) extended the study towards
southern declinations, observing a complete sample of WR stars
closer than 3 kpc. Up to now, only a few southern OB stars were
observed (e.g. Leitherer, Chapman \& Koribalski 1995). We have
initiated a programme with the aim of detecting bright early-type
stars that remain unobserved at radio wavelengths in the southern
sky. Here we present the results obtained so far, after the first
two observing campaigns.

The next section gives characteristics of the selected sample of
OB stars. Section~3 describes the observations and reduction
procedure. The theoretical formulae used here are briefly
described in Section~4. The results are presented in Section~5 and
discussed in Section~6. We close with the conclusions in
Section~7.

\section{The target stars}

From the lists published by Cruz-Gonz\'alez et al. (1974) and
Garmany et al. (1982) we selected all southern stars (declinations
south of --24$^{\circ}$), closer than 3 kpc and still unobserved
at cm wavelengths. These stars form our working database.

We derived an {\sl expected} mass loss rate with the approximation
given by Lamers \& Leitherer (1993). Their formula gives $\dot{M}$
as a function of effective temperature, $T_{\rm eff}$, and stellar
luminosity, $L_{*}$. The values for $T_{\rm eff}$ and $L_{*}$ were
taken from Vacca et al. (1996). By means of the standard model for
thermal emission from stellar winds (Wright \& Barlow 1975,
Panagia \& Felli 1975) we computed a minimum detectable mass loss
rate, assuming the radio flux as two times the predicted r.m.s.
after 3 hours integration time  with the ATCA. The predicted
r.m.s. noise resulted in 0.050 mJy at 3 cm, and 0.054 mJy at 6 cm,
if continuum emission from the target stars was not contaminated
with confusion from other sources in the field. Finally, we
defined an index $F$ equal to the expected mass loss rate over the
minimum detectable mass loss rate, and ranked our stars according
to that index.

The present target stars were chosen from the database depending
on the allocated time, and favoring the objects with highest index
$F$, i.e. the closest  and earliest. The candidates are listed in
Table~1. For this sample, the minimum detectable mass loss rates
were (1 -- 3) $10^{-6}$ M$_{\odot}$ yr$^{-1}$.

 We have included the already
observed star HD 57060 (29 CMa), which was not detected with the
VLA (Abbott et al. 1980), in the target list because, besides
matching most of the criteria, it is a very interesting Of+O
binary system. The r.m.s. noise expected from the ATCA, 0.050 mJy,
is 8$\sigma$ lower than the reported VLA flux limit of 0.4 mJy.
The masses of both the primary and secondary stars are in the
range 20--30 $M_{\odot}$, the primary being the more evolved star
(Hutchings 1977, Stickland 1989). Wiggs \& Gies (1993) interpreted
the H$\alpha$ P Cygni profile by means of a model where emission
comes from inside the wind of the primary and from a plane midway
between the stars where the winds collide.

H$\alpha$ emission has also been reported in HD 163181 (Thaller
1997) and thus this binary star can be suspected to have colliding
winds.

VLA 6-cm observations towards the field of HDE 319718 (N35
according to Neckel 1984) were presented by Felli et al. (1990).
Again, using the ATCA at 3 and 6 cm we intended to obtain a more
sensitive image.

Table~1 lists the stellar parameters, beginning with the name,
spectral classification and photometric data.  Information
concerning possible membership to a stellar association is given
in column 6. The spectro-photometric distance was estimated using
the $M_{\rm v}$ and $(B-V)_0$ values from Vacca et al. (1996) and
Schmidt-Kaler (1982), respectively. In the case of HD 163181, the
absolute magnitude was extrapolated. In all cases but HD 101205,
the distances estimated by us were, within errors, in agreement
with the distances to the OB associations. We adopted as the
stellar distances $d$ the cluster distances whenever available,
and the spectro-photometric distances for field stars. The mass
loss rates quoted in Table~1 were derived by means of optical and
UV spectra. When no measured terminal velocities ($v_\infty$) were
found in the literature, the values were interpolated from Prinja
et al. (1990) and appear in brackets. Column 9 accounts for the
binary status. The label ``BIN'' stands for the stars catalogued
as binaries with computed orbits. Following Gies (1987) SB2? means
that double lines have been reported, but no orbit was determined;
SB1? indicates that the star is a possible spectroscopic binary;
and C, that the radial velocity was taken as constant, i.e. as for
single stars.

\begin{table*}[h]
\begin{center}
\begin{tabular}{l c c c c c c c l}
\multicolumn{9}{l} {{\bf Table 1}.  Stellar parameters for the
target stars}\cr &&&&&&&&\\ \hline &&&&&&&&\\ Name & Sp.Class. &
$m_{\rm v}$ & {\it (B--V)} & $d$ & Assoc. & $\log(\dot{M})$ &
$v_\infty$ & Binary\\
 & & & & (kpc) & & (M$_{\odot}$/yr) & (km/s) & status \cr
\hline
&&&&&&&&\cr
HD 57060 & O8.5If$^1$ & 4.90$^1$ & --0.14$^1$ &
1.5 & NGC 2362$^2$ & --5.64$^3$ & 1425$^4$,1800$^5$ & BIN$^6$\\
 & O8Iaf+O$^7$ &  &  & & & --5.07$^8$ & 1700$^9$ \\
CD--47$^{\circ}$ 4551 & O5If$^{10}$ & 8.39$^{11}$ & 0.89$^{11}$ &
1.7 & &&[1885]  &
\\ HD 94963  & O6.5III(f)$^{12}$ & 7.18$^1$ &
--0.09$^1$ & 2.2& Car OB2$^2$ & &[2545]  & C$^2$\\ HD 97253  &
O5.5III(f)$^{12}$ & 7.12$^1$ & 0.15$^1$ & 2.2 & Car OB2$^2$ & &
[2600] & BIN$^2$ \\ HD 101205 & O7IIIn((f))$^{12}$ & 6.48$^1$ &
0.07$^1$ & 2.5 & Cru OB1$^2$ & & 2740$^4$ & SB2?$^2$ \\ HD 112244
& O8.5Iab(f)$^{12}$ & 5.33$^1$ & 0.01$^1$ & 1.5 &  & --5.30$^8$ &
1950$^6$,1880$^{13}$ & SB1?$^2$\\
 & & & & & & & 2160$^{14}$,1575$^{4}$&\cr
HD 124314 & O6V(n)((f))$^{12}$ & 6.64$^1$ & 0.21$^1$ & 1.0 & &
--5.15$^8$& 2500$^4$ & SB1?$^2$ \\ HD 135240 & O7.5III((f))$^{15}$
& 5.08$^1$ & --0.06$^1$ & 1.0 & & --6.21$^{16}$ &
2460$^4$,2700$^5$
 & BIN$^2$ \\
 HD 135591 & O7.5III((f))$^{15}$ & 5.43$^1$ & --0.09$^1$ & 1.2 & &
--6.40$^8$,--7.26$^{17}$ &  2600$^{17}$,2235$^4$ & BIN$^{18}$  \\
HD 150135 & O6.5V((f))$^{15}$ & 6.89$^1$ & 0.13$^1$ & 1.4 & Ara
OB1a$^2$ &  & [2455] & SB1?$^{2,6}$ \\ HD 150136 &
O5IIIn(f)$^{15}$ &5.64$^1$ & 0.14$^1$ & 1.4 & Ara OB1a$^2$ &  &
3160$^4$ & BIN$^6$
\\ HD 163181 & B1Iape$^{10}$ & 6.43$^{19}$ & 0.52$^{19}$ & 1.4
 & & --5.15$^8$ & 520$^4$ & BIN$^{20}$ \\
   & BN0.5Iap$^{21}$ \\
HDE 319718 & O7III$^{22}$,O3III$^{13}$  & 10.43$^{22}$ &
1.48$^{22}$ & 1.7& NGC 6357$^{23,24}$ &&[2295]  &  \\ &&&&&&&&\cr
\hline \multicolumn{9}{l} {1 -- Cruz-Gonz\'alez et al. (1974); 2
-- Gies (1987); 3 -- Drechsel et al. (1980); 4 -- Howarth et al.
(1997);}\cr \multicolumn{9}{l} {5 -- Hutchings \& von Rudloff
(1980); 6 -- Chlebowski \& Garmany (1991); 7 -- Garmany, Conti \&
Massey (1980);} \cr \multicolumn{9}{l}{8 -- Hutchings (1977); 9 --
Hutchings (1976); 10 -- Garrison et al. (1977); 11 -- Drilling
(1991); 12 -- Walborn (1973);}\cr \multicolumn{9}{l} {13 --
Vijapurkar \& Drilling (1993); 14 -- Bernabeu et al. (1989); 15 --
Walborn (1972a); 16 -- Prinja \& Howarth (1986);}\cr
\multicolumn{9}{l} {17 -- Conti \& Garmany (1980); 18 -- Lindroos
(1985); 19 -- Garmany, Conti \& Chiosi (1982); 20 -- Levato et al.
(1988);}\cr \multicolumn{9}{l} {21 -- Walborn (1972b); 22 --
Neckel (1984); 23 -- Johnson (1973); 24 -- Felli et al. (1990)}\cr
\end{tabular}
\end{center}
\end{table*}

\section{Observations and analysis}

The observations presented here were obtained with the Australia
Telescope Compact Array in two campaigns: 1) in 1998, February
16/17, using the 6A configuration, and 2) in 2000, March 30 +
April 1/2, with the 6D configuration, collecting a total of 52 h
observing time. In both sessions the maximum baseline was 6 km.

Each source was observed simultaneously at two frequencies: 8.64
GHz (3 cm) and 4.8 GHz (6 cm), with a total bandwidth each of 128
MHz over 32 channels. The average time on source was 2.5 to 3 h,
exception made for HD 124314 which was observed on three
occasions, in looking for flux variability. The stars were
observed during intervals of 15 or 30 minutes, interleaved with
short observations of close phase calibrators, and spanned over an
LST range of $\sim$11 hours for most of them. The allocated time
allowed us to track the sources HD 57060, HD 163181 and HDE 319718
for $\sim$7 h each. The flux density scale was calibrated using
observations of the primary calibrator, 1934--638, assuming flux
densities of 2.84 Jy at 3 cm and 5.83 Jy at 6 cm.

The data were reduced with Miriad routines and analyzed with the
AIPS package. After data calibration the visibilities were
Fourier-transformed using `natural-' and `uniform'-weighting. The
images obtained in the former way turned out to have better signal
to noise ratio. The restoring beams were $\ga$1\farcs5 at 3 cm and
$\ga$3\arcsec\ at 6 cm. In many cases, the diffuse emission from
extended sources towards the observed fields had to be removed by
taking out the visibilities corresponding to the shortest
baselines.

Flux density values of the thirteen observed sources are listed in
Tables~2a and 2b. The three entries of HD 124314 correspond to the
different observing dates (C1: Campaign 1, and C2: Campaign 2).
The optical positions were taken from the SIMBAD database. Four
sources were detected at both frequencies, with flux densities
greater than 3$\sigma$. HD~150135 was detected at 3 cm only; at 6
cm it appears at a $2.5\sigma$ level. All detected sources were
unresolved: CD--47$^{\circ}$ 4551, HD 124314, HD 150135, HD 150136
and HD 163181. Bi-dimensional Gaussians were fitted to the five
sources using Miriad task {\sl Imfit}, and the positions of the
maxima are given as the radio position in Table~2a. The
differences with optical coordinates, whenever present, are $\leq$
0\farcs5, so the optical and radio positions are in good agreement
within the errors. The flux density was computed after correcting
the  zero level  in the images. The flux density error in the
Tables is equal to the image r.m.s. noise ($\sigma$). For the
undetected stars, a maximum flux density of 3$\sigma$ is quoted.
The integration time on source is listed in Tables~2a and 2b as
$t_{\rm int}$.

Figures 1 to 4 display the detected stars, at 3 and 6 cm. The
restoring beam is plotted in each case.

\begin{figure}
 \resizebox{\hsize}{!}{\includegraphics{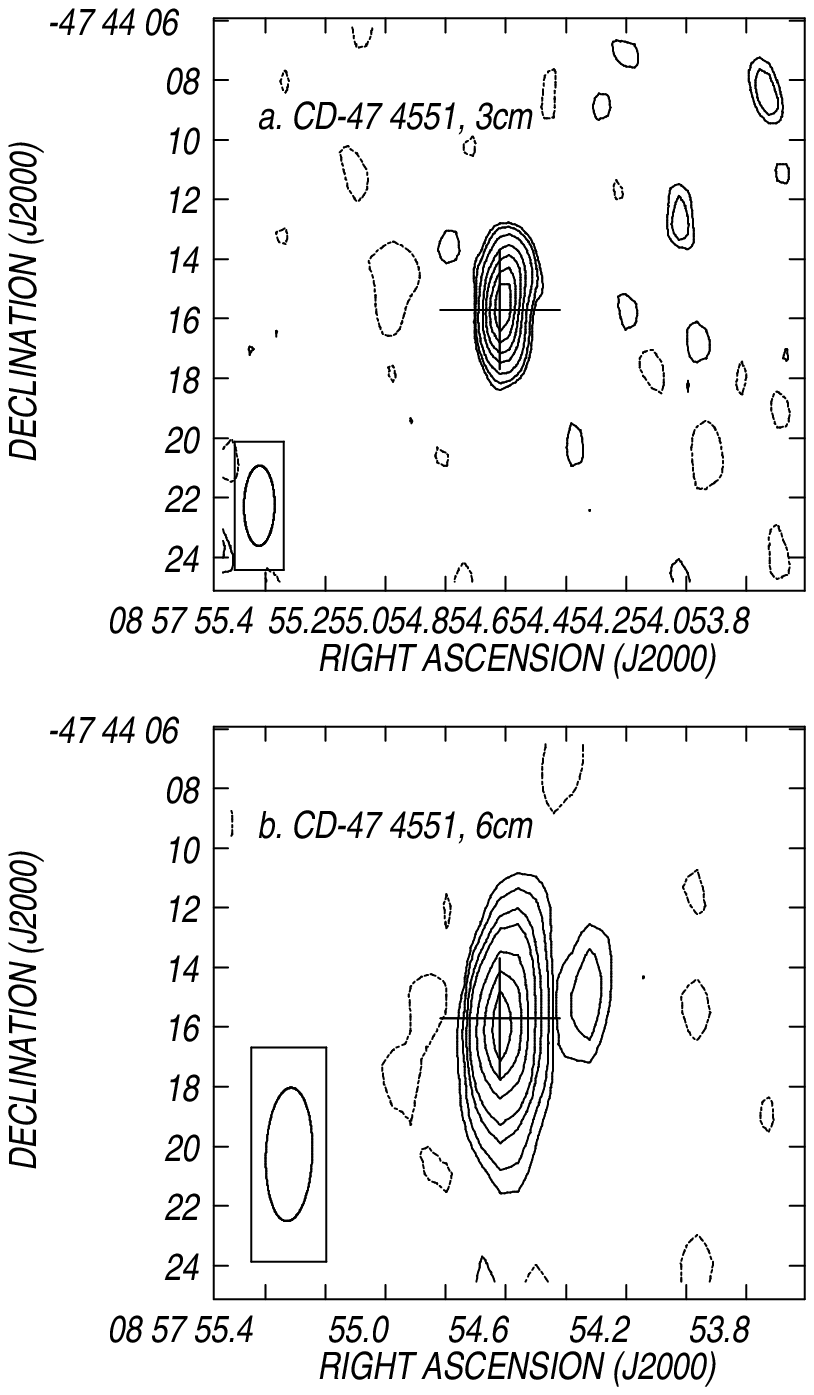}}
 \caption{{\bf a)} Contour map of CD--47$^{\circ}\,$4551 at 3
cm.  Contour levels are -0.1, 0.12(2.5$\sigma$), 0.18, 0.3, 0.5,
0.7, 0.95 and 1.2 mJy per beam area. The restoring beam, plotted
at the bottom left corner, is 2.7x1.0 arcsec in P.A. -2$^{\circ}$.
{\bf b)} The same as a), at 6 cm. Contour levels are --0.1,
0.18(3.5$\sigma$), 0.3, 0.5, 0.7, 1.2, 1.7 and 2.2 mJy per beam
area. The restoring beam is 4.5x1.6 arcsec in P.A. -2$^{\circ}$.
Negative contours are dashed. The optical position of the star is
marked by a cross.}
  \label{fig1} \end{figure}

\begin{figure}
 \resizebox{\hsize}{!}{\includegraphics{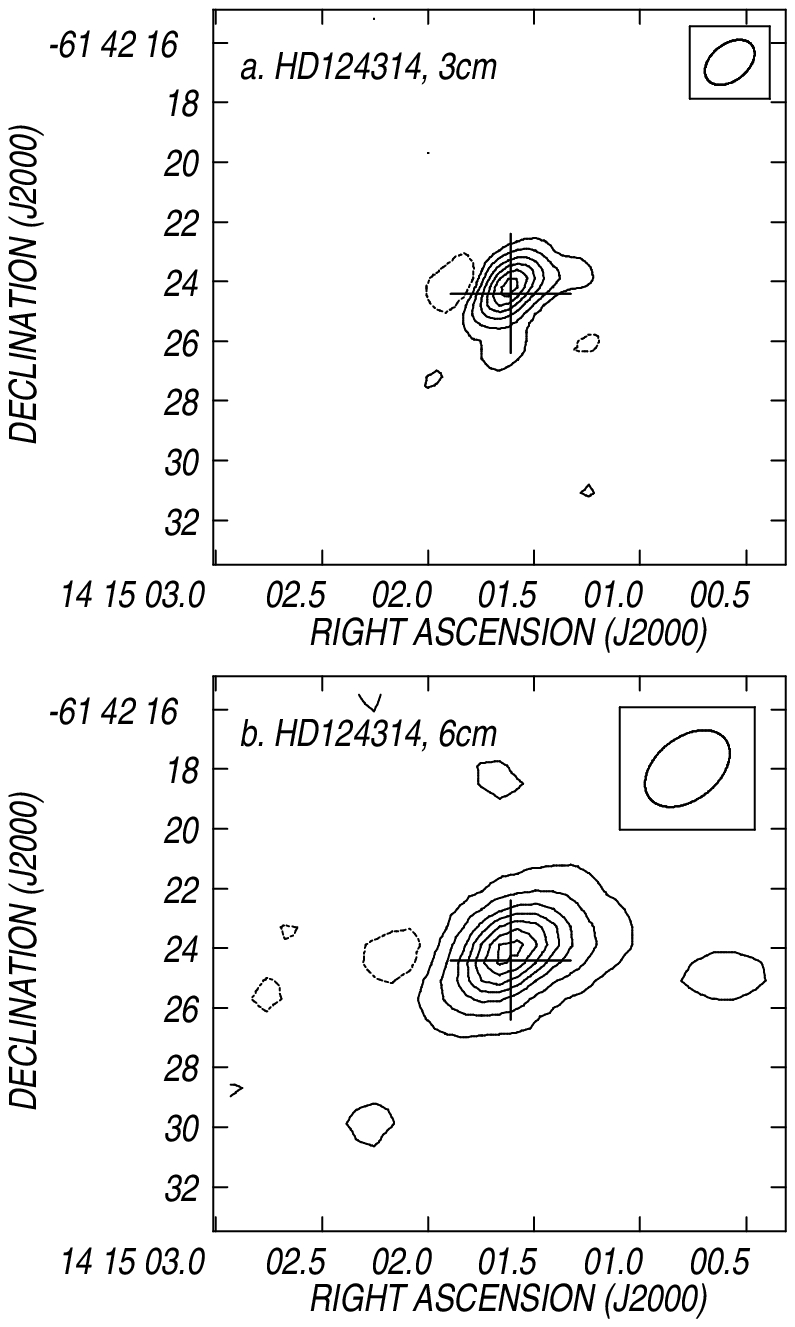}}
 \caption{{\bf a)} Contour map of HD 124314 at 3 cm. Contour levels are
--0.15, 0.15(3$\sigma$), 0.5, 1.0, 1.5, 2.0 and 2.5 mJy per beam
area. The restoring beam, plotted at the upper right corner, is
1.9x1.2 arcsec in P.A. -52$^{\circ}$. {\bf b)} The same as a) at 6
cm. Contour levels are --0.15, 0.15(3$\sigma$), 0.6, 1.2, 1.7,
2.2, 2.7 and 3.2 mJy per beam area. The restoring beam is 3.2x2.1
arcsec in P.A. -52$^{\circ}$.}
  \label{fig2} \end{figure}

\begin{figure*}
 \resizebox{\hsize}{!}{\includegraphics{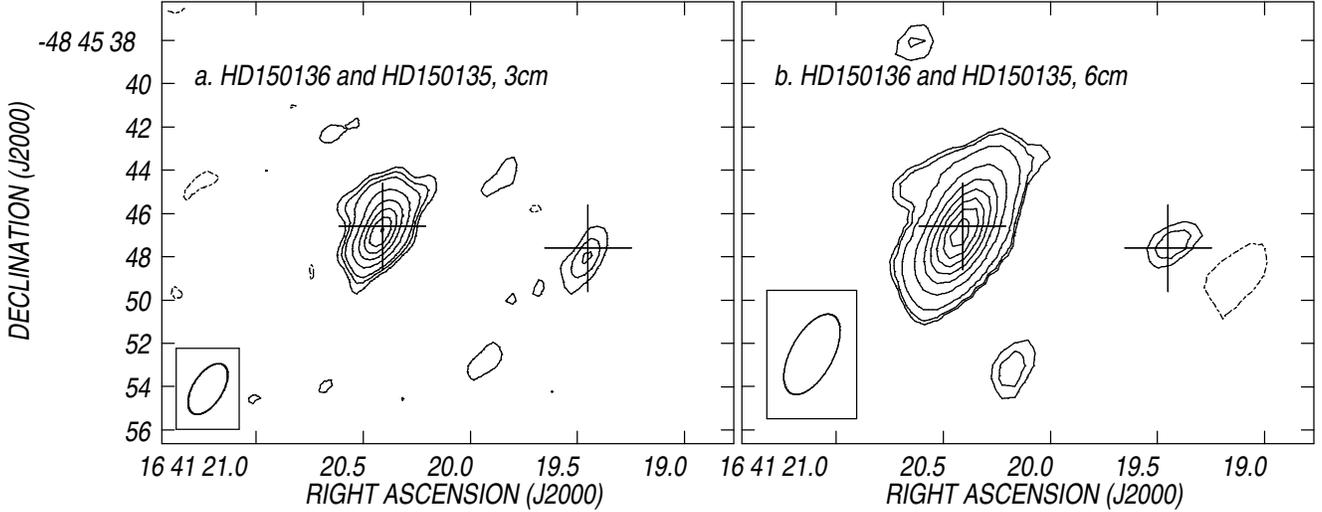}}
 \caption{{\bf a)} Contour map of HD 150135 and HD 150136 at 3 cm.
Contour levels are --0.08, 0.12(4$\sigma$), 0.2, 0.3, 0.6, 1.0,
1.4, 1.8 and 2.2 mJy per beam area. The restoring beam, plotted at
the bottom left corner, is 2.6x1.4 arcsec in P.A. -33$^{\circ}$.
{\bf b)} The same as a) at 6 cm. Contour levels are --0.15,
0.15(5$\sigma$), 0.2, 0.4, 1.0, 1.75, 2.5, 3.25, 4.0 and 4.75 mJy
per beam area. The restoring beam is 4.1x2.0 arcsec in P.A.
-33$^{\circ}$.}
  \label{fig3} \end{figure*}

\begin{figure}
 \resizebox{\hsize}{!}{\includegraphics{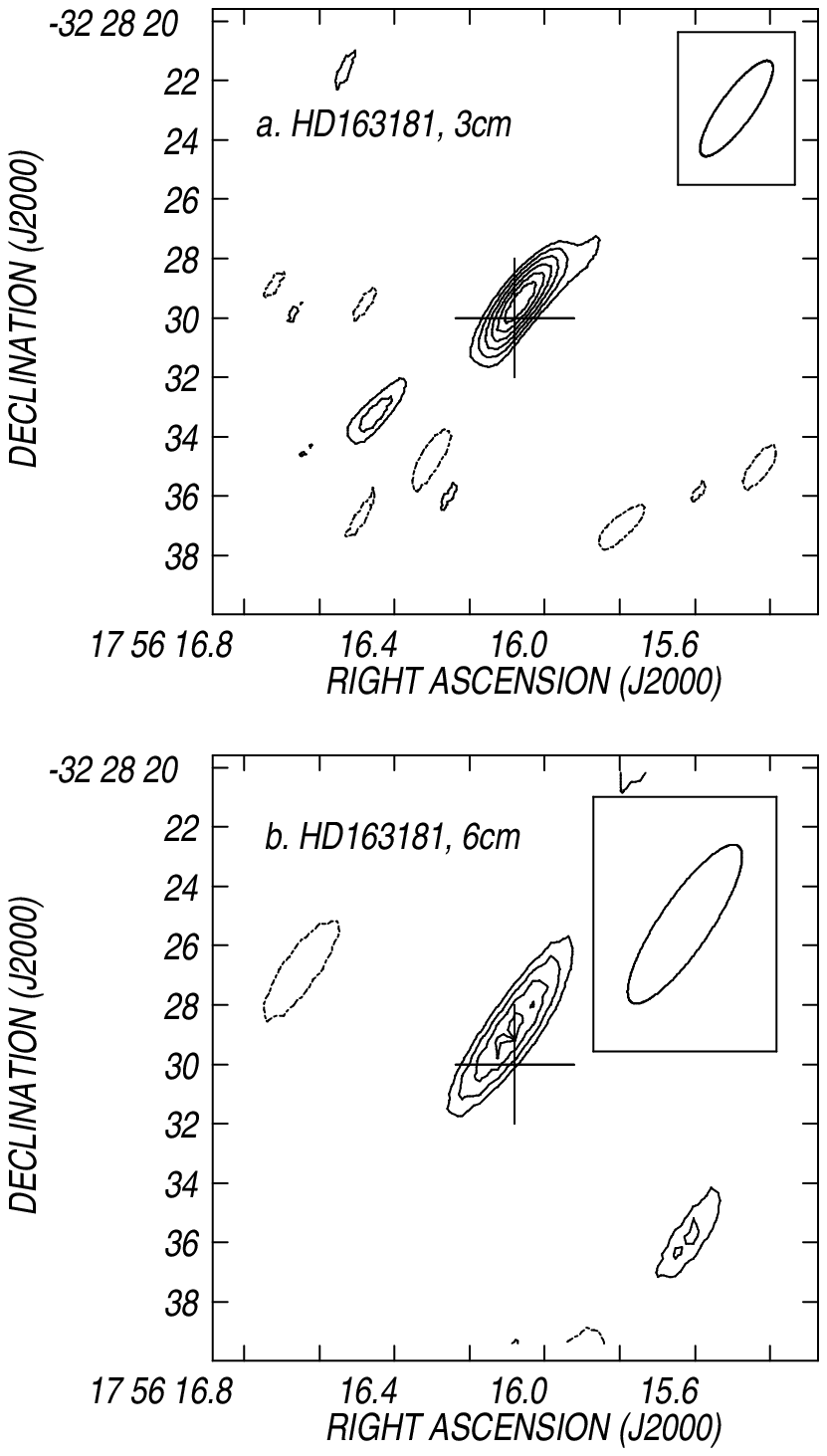}}
 \caption{{\bf a)} Contour map of HD 163181 at 3 cm.  Contour levels are
--0.15, 0.15(3$\sigma$), 0.2, 0.25, 0.3, 0.35 and 0.4 mJy per beam
area. The restoring beam, plotted at the upper right corner, is
3.9x1.1 arcsec in P.A. -34$^{\circ}$. {\bf b)} The same as a) at 6
cm. Contour levels are --0.13, 0.13(2.5$\sigma$), 0.17, 0.21 and
0.25 mJy per beam area. The restoring beam is 6.4x1.7 arcsec in
P.A. -34$^{\circ}$.}
  \label{fig4} \end{figure}

\begin{table*}
\begin{center}
\begin{tabular}{l l c c c c c c c}
\multicolumn{8}{l} {{\bf Table 2a}.  Positions and flux densities
of the detected stars}\cr &&&&&&&\cr \hline &&&&&&&\cr Star &
$\,\,\,\,\,\,\,\,$Optical & positions$\,\,\,\,\,$ &
$\,\,\,\,\,\,\,\,$Radio & positions$\,\,\,\,\,$ & $t_{\rm int}$ &
$S_{\rm 3 cm}$ & $S_{\rm 6 cm}$ \cr & R.A.(2000) & Dec.(2000) &
R.A.(2000)& Dec.(2000) & (min) &
 (mJy) &  (mJy) \cr
\hline &&&&&&&\cr

CD--47$^{\circ}$ 4551 & 08 57 54.62 & --47 44 15.7 & 08 57 54.61 &
--47 44 15.7 & 160 &  1.77$\pm$0.05 & 2.98$\pm0.05$\cr &&&&&&&\cr

HD 124314 & 14 15 01.61 & --61 42 24.4 & 14 15 01.61 & --61 42
24.2 &&\cr

$\,\,\,\,$C1: 02/16/1998 &  & & & & 150 & 1.88$\pm$0.05 &
2.72$\pm$0.08\cr

$\,\,\,\,$C2: 03/30/2000& & & & & 195 & 2.71$\pm$0.08 &
4.50$\pm$0.07\cr

$\,\,\,\,$C2: 04/02/2000& & & & & 130 & 2.88$\pm$0.05 &
3.78$\pm$0.05\cr

&&&&&&&\cr HD 150135 & 16 41 19.45 & --48 45 47.6 & 16 41 19.45 &
--48 45 47.8 & 185 & 0.28$\pm$0.03 & 0.10$\pm$0.04 \cr

HD 150136 & 16 41 20.41 & --48 45 46.6 & 16 41 20.42 & --48 45
46.8 & 185 & 2.61$\pm$0.03 & 5.57$\pm$0.03\cr

HD 163181 & 17 56 16.08 & --32 28 30.0 & 17 56 16.07 & --32 28
29.5 & 135 & 0.44$\pm$0.05 & 0.20$\pm$0.05 \cr&&&&&&&\cr \hline
\end{tabular}
\end{center}
\end{table*}

\begin{table*}
\begin{center}
\begin{tabular}{l c c c c c}
\multicolumn{6}{l} {{\bf Table 2b}.  Flux density limits of the
undetected stars}\cr &&&&&\cr \hline &&&&&\cr Star &
$\,\,\,\,\,\,$Optical &positions$\,\,\,\,\,$ & $t_{\rm int}$ &
$S_{\rm 3 cm}$ & $S_{\rm6 cm}$ \cr
 & R.A.(2000) & Dec.(2000) & (min)& (mJy) &  (mJy) \cr
\hline &&&&&\cr
 HD   57060 & 07 18 40.38 & --24 33 31.3 & 180 &
$<$0.15 & $<$0.15 \cr
HD   94963 & 10 56 35.79 & --61 42 32.4 &
140 & $<$0.14 & $<$0.17 \cr
HD   97253 & 11 10 42.02 & --60 23
04.3 & 140 & $<$0.15 & $<$0.15 \cr
HD  101205 & 11 38 20.36 & --63
22 21.9 & 145 & $<$0.13 & $<$0.20 \cr
HD  112244 & 15 55 57.13 &
--56 50 08.9 & 140 & $<$0.13 & $<$0.15 \cr
HD  135240 & 15 16
56.90 & --60 57 26.1 & 170 & $<$0.13 & $<$0.16 \cr
HD  135591 & 15
18 49.14 & --60 29 46.8 & 180 &$<$0.15 & $<$0.15 \cr

HDE 319718 & 17 24 42.90 & --34 11 48.0 & 180 & $<$0.15 & $<$0.40
\cr &&&&&\cr
\hline
\end{tabular}
\end{center}
\end{table*}

\section{Radio emission from stellar winds}

Stars with ionized winds show a flux density excess towards long
wavelengths. Assuming the continuum radiation is due to free-free
emission (thermal bremsstrahlung), Wright \& Barlow (1975) and
Panagia \& Felli (1975) deduced the spectrum in the radio and IR
energy  ranges. Their model was developed for a uniform flow
expanding at constant velocity, and allows to derive the stellar
mass loss rate as a function of the measured flux density as:

\begin{equation}
 \dot{M} = 5.32 \times 10^{-4} \, \frac{S_\nu^{3/4}\,d^{3/2} \,
v_\infty \, \mu}{Z\, \sqrt{\gamma\, g_\nu \,\nu}} \,\,\,\,\,\,
M_{\odot} yr^{-1}.$$
\end{equation}

Here $d$ is the stellar distance in kpc, $v_\infty$ the wind
terminal velocity in km s$^{-1}$, $S_\nu$ the flux density in mJy; $\mu$
stands for the mean molecular weight, $\gamma$ is the mean number
of electrons per ion  and $Z$, the rms ionic charge.

In the range of the wind electron temperatures $T_{\rm e}$ and observing
frequencies we are dealing with, the free-free Gaunt factor can be
approximated by:

\begin{eqnarray}
 g_\nu = 9.77\, \left( 1 + 0.13 \log \frac{T_e^{3/2}}{Z\nu}
\right) . \nonumber
\end{eqnarray}

The model predicts a spectral index $\alpha$ = 0.6 ($S_\nu \propto
\nu^\alpha$) in accordance with measurements of various WR and O
stars. If the spectral index differs from 0.6, Eq. (1) yields an
upper limit to the stellar mass loss rate. However, when detected
at two or more radio continuum frequencies, an important number of
stars show spectral indices approaching zero and even negative
values. The interpretation of these results is consistent with a
picture where the observed flux density has an important
non-thermal contribution. The main non-thermal process which is
relevant for early type stars is synchrotron radiation by
relativistic electrons, accelerated by first order Fermi mechanism
at strong shock fronts. The presence of non-thermal radio emission
provides indirect evidence for the presence of magnetic fields. In
some binary stellar systems the electrons seem to be accelerated
at the wind collision region (i.e. Cyg OB2 No. 5 and WR 140,
Contreras et al. 1997, 1996). Eichler \& Usov (1993) have shown
that stellar winds collision in early type binaries may be strong
sources of non-thermal radio and $\gamma$-ray emission (see also
Benaglia et al. 2001 for an updated discussion).

For single stars, White (1985) has shown that electrons can be
accelerated up to relativistic energies at the multiple shocks
formed at the base of the wind by line-driven instabilities. The
derived radio spectrum follows an index of $\sim$0.5 up to a
turnover around a few GHz. The relativistic electron population
would generate inverse-Compton $\gamma$-rays too. Benaglia et al.
(2001) have estimated the expected $\gamma$-ray luminosity from
the wind colliding region, the terminal shock of the wind and the
zone of the base of the wind, for Cyg OB2 No. 5, using the radio
results of Contreras et al. (1997).

An interesting feature seen in some non-thermal sources is their
variations in radio luminosity with time. Abbott et al. (1984)
found this behaviour in 9\,Sgr and Cyg OB2 No.\,9. Cyg  OB2 No.\,5
has been extensively monitored, since $\sim$1980 (see Contreras et
al. 1997 and references therein). The radio source seems to switch
between states of high and low emission with a period of 7 years.

\section{Results}

We detected five stars out of thirteen: CD--47$^{\circ}$ 4551, HD
124314, HD 150135, HD 150136 and HD 163181. Table~3 presents the
spectral indices. The radio emission of the candidates
CD--47$^{\circ}\,$4551, HD 124314 and HD 150136 is characterized
by a negative spectral index between 4800 and 8640 MHz. The one of
HD 124314 displays strong variations. The flux densities of these
stars seem arising mainly of a non-thermal process. The only B
star in the group, HD 163181, and HD 150135 show positive spectral
indices, possibly  indicating thermal emission as dominant.

\begin{table}
\begin{center}
\begin{tabular}{l r}
\multicolumn{2}{l} {{\bf Table 3}. Spectral indices of the
detected stars}\cr &\cr \hline &\cr Star & Spectral index \cr
 & $\alpha_{\rm 3-6cm}$ \cr \hline
&\cr CD--47$^{\circ}\,$4551 &  $-0.89\pm0.06$  \cr HD 124314 & \cr
$\,\,\,\,\,\,$ C1: 02/16/1998 & $-0.60\pm0.07$ \cr $\,\,\,\,\,\,$
C2: 03/30/2000 & $-0.87\pm0.06$ \cr $\,\,\,\,\,\,$ C2: 04/02/2000
& $-0.46\pm0.04$ \cr HD 150135   &   $+1.75\pm0.70$  \cr HD 150136
&   $-1.29\pm0.03$  \cr HD 163181   &   $+1.34\pm0.47$  \cr &\cr
\hline
\end{tabular}
\end{center}
\end{table}

From hereafter we shall call ``thermal'' sources the ones with
positive spectral index, and ``non-thermal'' if the index is
negative.

The measured flux densities at 3 cm were used to derive the mass
loss rates of all target stars, because of their smaller errors
compared to the flux densities at 6 cm. According to their
evolutionary status, for most of the target O-type stars the
computed mean molecular weight was 1.5. We adopted a value of 1.3
for HD 124314 and HD 150135, the less evolved stars, and 1.6 for
the only B-type star in the group. In deriving the rms ionic
charge and the mean number of electrons per ion we assumed that
both Hydrogen and Helium were singly ionized, and that in the
region where the radio emission comes from, the He$^{++}$ had the
chance to recombine. The electron temperature was assumed as 0.4
the effective stellar temperature. The computed values of $\mu$,
$Z$, and $\gamma$ are listed in Table~4, together with the
free-free Gaunt factor at 3 cm.

We derived the mass loss rates by means of Eq. (1), independently
of the behaviour of the flux density with frequency (see Table~4).
The results for candidates with spectral index $\alpha < 0$ appear
in parenthesis. For non-detections and non-thermal sources, the
derived mass loss rates must be considered as upper limits.

In the computation of the mass loss rate errors we have adopted
the relative deviations for $\mu$, $\gamma$, $g_\nu$ and $Z$ of
$\sim$10\%, and for the terminal velocity an accuracy better than
20\%. The flux density errors contribute with $\sim$10\% in the
worst cases. Undoubtedly the main source of error in $\dot{M}$ is
the uncertainty in stellar distances. If the relative error is of
20\%, the resulting $\sigma(\log(\dot{M})) \approx 0.2$. This last
value increases to 0.3 in the more conservative case of distances
accurate up to 30\%.

Inhomogeneities or anisotropy in the winds like clumpiness or
collimated flows are other causes of errors in $\dot{M}$ when
estimated from the model of Wright \& Barlow (1975).

 Among the eight
non-detected stars, five had predicted mass loss rates of the
order of the minimun detectable mass loss rates, and we expected
that they would be detected at a 2$\sigma$ level. The stars HD
57060 and HD 112244 have both expected mass loss rates, and mass
loss rates derived by optical means appreciably larger than the
minimum detectable mass loss rates.

\begin{table*}
\begin{center}
\begin{tabular}{l c c c c c r}
\multicolumn{7}{l} {{\bf Table 4}. Radio-derived mass loss rates
}\cr &&&&&&\\ \hline &&&&&&\cr Star & $T_{\rm e}$ & $\mu$ & $Z$ &
$\gamma$ & $g_{\rm 3cm}$ & $\log(\dot{M})$ \cr
 & (K) &  &&  & &  (M$_{\odot}$/yr)\cr \hline
&&&&&&\cr HD 57060 & 14300 & 1.5 & 1.0 & 1.0 & 5.1 &  $<-5.52$ \cr
CD--47$^{\circ}\,$4551 & 17900  & 1.5 & 1.0 & 1.0 & 5.3 &
($-$4.62) \cr  HD 94963 & 16500 & 1.5 & 1.0 & 1.0 & 5.2 &
$<-5.15$ \cr

HD 97253 & 17600 & 1.5 & 1.0 & 1.0 & 5.2 &  $<-5.11$  \cr

HD 101205 & 16000& 1.5 & 1.0 & 1.0 & 5.2 &  $<-5.08$  \cr

HD 112244 & 13700& 1.5 & 1.0 & 1.0 & 5.0 & $<-5.54$  \cr

HD 124314 & 17400& 1.3 & 1.0 & 1.0 & 5.2 & ($-$4.74) \cr

HD 135240 & 15400& 1.5 & 1.0 & 1.0 & 5.1 & $<-5.69$   \cr

HD 135591 & 15400& 1.5 & 1.0 & 1.0 & 5.1 & $<-5.49$ \cr

HD 150135 & 16900& 1.5 & 1.0 & 1.0 & 5.2 & $-$5.23 \cr

HD 150136 & 18200& 1.5 & 1.0 & 1.0 & 5.3 & ($-$4.40)   \cr

HD 163181 & 11600& 1.6 & 0.9 & 0.8 & 5.0 & $-$5.63   \cr

HDE 319718 & 15900& 1.5 & 1.0 & 1.0 & 5.2 & $<-5.34$   \cr
&&&&&&\cr \hline
\end{tabular}
\end{center}
\end{table*}

\subsection{Comments on individual stars}

{\sl HD 57060}. This source was observed for 3 hours with the ATCA
and was not detected. The r.m.s. of the final images agrees with
the expected r.m.s. of 0.05 mJy at 3 cm, and 0.054 mJy at 6 cm.
According to these values we conclude that the flux density of
this star must be below 0.15 mJy, and the mass loss rate less than
3 $\times$ 10$^{-6}$ M$_{\odot}$ yr$^{-1}$. Abbott et al. (1980)
reported no detection at 6 cm above 0.4 mJy after 165 minutes of
integration time with the VLA.\\

{\sl CD--47$^{\circ}\,$4551}. Besides the target star, at 4800 MHz
a second source appears 5.4\arcsec\ to the west, with a maximum
over 6$\sigma$ (Fig. 1b). Its flux density is 0.33 mJy at 6 cm.
The non detection of this small source at 3 cm means that the flux
density is $S_{3 \rm cm} < 0.15$ mJy, letting us to conclude that
the spectral index is negative. At 1.7 kpc the spatial distance
between the two maxima is about 9000 AU.\\

{\sl HD 94963, HD 97253, HD 101205, HD 112244, HD 135240 and HD
135591}. The observations allowed us to estimate upper limits to
mass loss rates for these undetected stars.\\

{\sl HD 124314}. This star was first detected on February 1998.
The measured flux densities at 3 and 6 cm yield a non-thermal
spectral index of $-0.60$. Fig. 2 depicts the stellar wind as
observed the first day of C2. In order to look for variability
with timescales of years and days, we reobserved the source on
March 30 and April 2, 2000.

For this particular star we produced, at each frequency, three
images, maintaining the number of iterations, the baselines
(greater than 10 k$\lambda$) and the restoring beams (2\arcsec\ at
3 cm and 3.6\arcsec\ at 6 cm). Although we were not able to
appreciate differences in shape between the maps made using data
taken at different dates, the observed flux densities showed large
variations (see Table~2a).

At 3 cm the variability is appreciable between observations of
1998 and 2000, as an increase of $\sim$50\%. Flux densities of C2
did not present differences within the error.

The source displayed always variations when observed at 6 cm. The
flux densities measured at 2000 are 65\% and 40\% greater than the
corresponding one of 1998. During the three days-interval of C2
the flux decreased about 15\%.

The mass loss rate of HD 124314 was derived from the observations
with the highest spectral index, i.e., for which the non-thermal
contribution to the flux densities were the smallest. \\

{\sl HD 150135 and HD 150136}. Fig. 3 displays the radio images of
these two stars. HD 150135 is barely detected at 6 cm, with $S_{6
\rm cm} \approx 2.5\sigma$. The earliest and strongest source, HD 150136, has
a singular shape. Although elongated as the observing beam, it
shows extensions to both sides, visible at the two observing
frequencies.\\

{\sl HD 163181}. At 6 cm it is a 4$\sigma$ detection, being the
flux density about half of that at 3 cm (see Fig. 4 and
Table~2b).\\

{\sl HDE 319718}. This star is located near the H\,{\sc ii} region
G353.2$+$0.9. The 6-cm VLA map of the ionized gas (Felli et al.
1990) has a peak of $\sim$ 120 mJy, and covers a region of
2\arcmin\ in diameter, 1\arcmin\ north of our target star. To
eliminate completely the emission from the extended sources, the
ATCA image was built considering visibilities of baselines only
with antenna 6. The candidate was undetected above 0.15 mJy at 3
cm. However, upper limits to the flux density permitted us to
derive a corresponding limit to the mass loss rate.

\subsection{Other sources in the fields}

Towards the direction of the target stars two previously unknown
sources were detected, named S1 and S2. They are visible
at the two observed frequencies and are point sources for ATCA.

S1 is located in the field of HD 94963, at the radio position
$\alpha = 10^{\rm h} 56^{\rm m} 29.27^{\rm s}$, $\delta = -61^{\rm
o} 43' 1.19''$. The measured flux densities are $S_{\rm 3cm} =
3.8\pm0.05$ mJy and $S_{\rm 6cm} = 6.4\pm0.06$ mJy, thus the
spectral index results in --1.16.

The second source is located at $\alpha = 15^{\rm h} 18^{\rm m}
35.49^{\rm s}$, $\delta = -60^{\rm o} 29' 59.44''$ in the field of
HD 135591. Its radio flux densities are $S_{\rm 3cm} =
0.24\pm0.05$ mJy and $S_{\rm 6cm} = 0.60\pm0.05$ mJy, giving a
spectral index of --0.68.

Possible counterparts to these objects were searched using NED and
SIMBAD databases and SkyView\footnote{SkyView was developed under
NASA ADP Grant NAS5-32068 with Principal Investigator Thomas A.
McGlynn under the auspices of the High Energy Astrophysics Science
Archive Research Center (HEASARC) at the GSFC Laboratory for High
Energy Astrophysics.} images from the Digitized Sky
Survey\footnote{The Digitized Sky Survey was produced at the Space
Telescope Science Institute under US Government grant NAG W-2166.}
(DSS), with no positive results. The radio negative spectral
indices indicate a non-thermal origin of the emission. This fact,
together with the small angular size of the sources and the non
existence of other related object suggest an extragalactic origin
for S1 and S2.

\section{Discussion}

\subsection{Statistics on OB radio observations}

 In order to perform some statistics on radio
detections of OB stars, we have consulted the articles of Abbott
et al. (1980, 1981, 1984), Bieging et al. (1989), Leitherer et al.
(1995), Contreras et al. (1996), Waldron et al. (1998) and Scuderi
et al. (1998), and added the results of this work. Regarding these
investigations, about 120 OB stars have been observed, and less
than 40 detected at least at one frequency.

Excluding our observations, about 25 stars were detected at two or
more frequencies, allowing the determination of spectral indices.
Of the almost 40 detected sources, $\sim$50\% classify as thermal,
$\sim$25\% as non-thermal, and $\sim$25\% as composite or
undecided.

No non-thermal B stars were found, giving support to the
suggestion of Bieging et al. (1989) that the synchrotron mechanism
is less efficient at lower luminosities.

After the detections presented here the thermal candidates stay
around 50\% between the detected O and B-type stars, while the
non-thermal ones have increased from 25\% to 30\%. For detected
O-type stars, the figures are more striking: the thermal
candidates are $\sim$40\% (before and after this work), but the
percentage of non-thermal O stars changes from 40\% to almost 50\%
after the present results. Chapman et al. (1999) arrived at a
similar result for WR stars, in surveying all candidates at
negative declinations and distances less than 3 kpc. We think that
the number of non-thermal candidates must be taken as a lower
limit because:

\noindent {\sl i)} a single O star with powerful winds can radiate
non-thermal radio emission due to relativistic leptons. According
to the model of White (1985), a spectral index of $+0.5$ will be
measured at low frequencies, mimicking a thermal spectrum.

\noindent {\sl ii)} the flux density of a non-thermal source is
known to vary, and at the time of the observation a minimum can be
occurring, approaching a thermal spectrum.

\subsection{Flux variability and HD 124314}

It is important to analyze the variations of the flux density in
HD 124314. They could be originated in changes in the mass loss
rate, ionization or density structure that take place throughout
the photosphere of the star. The corresponding time scale depends
on the observing frequency. Scuderi et al. (1998) and Waldron et
al. (1998) give expressions to estimate the time scales of the
variations.

Let $R_{\rm eff}$ be the characteristic radius of the emitting
region, given by Wright \& Barlow (1975) for a thermal outflow.
When considering the mass loss rate predicted using Vink et al.
(2000) models for HD 124314, $R_{\rm eff} \approx$ 1200
$R_{\odot}$ and the transit time of the shock front in the radio
regime will be $>$ 5 days for variations in  $\dot{M}$. On the
other hand, variations in ionization are characterized by time
scales of $\geq$ 20 days in the radio emission. The flux density
changes must be faster towards higher frequencies, as $R_{\rm
eff}$ diminishes.

Following the model of Waldron et al. (1998), the  time travel for
a density disturbance to propagate from the X-ray radius to the
radio radius is above 10 days.

Taking the above considerations into account, is clear that the
flux variation between C1 and C2 can be ascribed to the mentioned
phenomena. In contrast, it seems difficult to find the same causes
for the differences between the flux densities of the two
observing runs during C2, which are separated just three days and
occur at 6 cm but not at 3 cm. A simpler explanation is to assume
that the quoted flux density errors are underestimated.
Nonetheless, it will be interesting to perform a series of
observations with time scales of a few days towards this star,
which is proposed as a counterpart to the unidentified gamma-ray
source 3EG J1410-6147 (Romero et al. 1999).

\subsection{Comparison between radio-derived $\dot{M}$ and other estimates}

The mass loss rates derived here were compared to observed values
found in the literature and from the new models of Vink et al.
(2000). For the target stars, previous estimates of $\dot{M}$ were
derived from optical and UV profiles. The theoretical -expected-
mass loss rates can be computed from Vink et al.'s models, as a
function of the effective temperature, luminosity, mass, and wind
terminal velocity. For our calculations we have adopted the
stellar parameters given by Vacca et al. (1996). One must bear in
mind that predicted values were calculated with parameters derived
under a series of assumptions from models, depending on a certain
spectral classification, which is not unique.

Table~5 lists the mass loss rates obtained from the literature,
Vink et al., and us. For the correlation we have assumed error
bars of $\sim$50\% $\dot{M}$ in the values derived in this work.

The result of the comparison showed no pattern. For the stars HD
94963, HD 97253, HD 101205, HD 135591 and HDE 319718 the predicted
and the radio values are in good agreement (see columns 3 and 4 of
Table~5), although we would have expected radio signal around a
2$\sigma$ level for most of them.

Despite HD 57060 spectral classification (Of+O), its predicted and
optically derived mass loss rate, it was not detected at 3 or 6
cm. Using Eq. (1) we computed an expected flux density, $S^*_\nu$,
if $\dot{M} = \dot{M}_{\rm optical} = 8.51 \times 10^{-6}$
M$_{\odot}$ yr$^{-1}$. We found $S^*_{\rm 3cm} \approx 0.6 {\rm
mJy}
> 4S_{\rm 3cm}$ and $S^*_{\rm 6cm} \approx 0.4 {\rm mJy} > 2S_{\rm
6cm}$. The same relation of predicted and optically derived with
radio mass loss rate happens to HD 112244. Here  $S^*_{\rm 3cm}
\approx 0.27 {\rm mJy} > 2S_{\rm 3cm}$.

It is difficult to believe that the discrepancies between mass
loss rates showed above might arise from underestimates in stellar
distances, values which are, for both stars, widely accepted.
Uncertainties in the wind terminal velocities are evident from
Table~1. Different methods of determining $v_\infty$ yield to a
range of values: $\sim$ 1400 -- 1800 km s$^{-1}$  for HD 57060,
and $\sim$ 1600 -- 2200 km s$^{-1}$ for HD 112244. In our
calculations we have adopted $v_\infty({\sc {\rm HD}\,57060}) =
1800$ km s$^{-1}$ and $v_\infty({\sc {\rm HD}\,112244}) =  1900$
km s$^{-1}$. But the error on $v_\infty$ does not suffice to
account for the non-detections at radio continuum. We speculate
that the presence of variability in the mass loss rates might be
an explanation. Other possibility is that some of the assumptions
undertaken in deriving $\dot{M}$ from optical lines (plus some
theoretical modeling) (Hutchings 1976) are not entirely valid.

For HD 135240, probably the expected mass loss rate computed using
Vink et al. (2000) formulae is an overestimation. The radio upper
limit is consistent with the mass loss rate from Hutchings (1976).

The stars CD--47$^{\circ}\,$4551, HD 124314 and HD 150136, with
negative spectral indices, have theoretical mass loss rates lower
than the values obtained from the radio observations, as expected.
The negative spectral indices are clear evidence of synchrotron
contamination at 3 and 6 cm, besides the free-free contribution.
This contamination may lead to the extreme case of an
overestimation in $\dot{M}$ of about an order of magnitude for HD
124314 and HD 150136, as Contreras et al. (1996) found for Cyg OB2
No.\,9. The mass loss rate determined by Hutchings (1976) is
similar to the radio one.

In the case of HD 150135, ($\alpha > 0$), the $\dot{M}$ derived
here resulted greater than the predicted one, as for the stars
with negative spectral index. In the model of White (1985)
non-thermal emission is capable of producing a spectrum with
$\alpha \sim 0.5$. The search for variability can help to
discriminate between a thermal or non-thermal emission mechanism.

Bieging et al. (1989) reported that the observed mass loss rates
 were less than predicted for B supergiants of high luminosity. They
 pointed out that one source of error was that the model used to compute the
 expected rates neglected multiply  scattered photons. Notwithstanding the model
 of Vink et al. (2000) to derive $\dot{M}$ takes into account scattered
 photons, the radio-derived mass loss rate of HD 163181 is less than predicted
in a factor of 2.

\begin{table}
\begin{center}
\begin{tabular}{l c c r}
\multicolumn{4}{l} {{\bf Table 5}.  Non-radio (n-r), predicted
(VKL), and radio- } \cr \multicolumn{4}{l} {derived mass loss
rates }\cr &&&\cr \hline &&&\cr Star & $\dot{M}$ (n-r) & $\dot{M}$
(VKL)$^1$ & $\dot{M}$ (radio)\cr & (M$_{\odot}$/yr) &
(M$_{\odot}$/yr) & (M$_{\odot}$/yr) \cr
 \hline
&&&\cr
HD 57060 & 8.51 $10^{-6}$ & 7.85 $10^{-6}$ & $<$ 3.03
$10^{-6}$ \cr
& 2.29 $10^{-6}$ & & \cr

CD--47$^{\circ}\,$4551 & & 1.81 $10^{-5}$ & 2.39 $10^{-5}$ \cr

HD 94963 & & 3.34 $10^{-6}$ & $<$ 7.16 $10^{-6}$  \cr

HD 97253 &  &5.22 $10^{-6}$& $<$ 7.70 $10^{-6}$ \cr

HD 101205  &  & 2.51 $10^{-6}$ & $<$ 8.38 $10^{-6}$ \cr

HD 112244 & 5.01 $10^{-6}$ & 4.06 $10^{-6}$ & $<$ 2.87 $10^{-6}$ \cr

HD 124314 & 7.08 $10^{-6}$  & 1.91 $10^{-6}$  & 1.81 $10^{-5}$ \cr

HD 135240 & 6.17 $10^{-7}$ & 2.02 $10^{-6}$  & $<$
2.03 $10^{-6}$ \cr

HD 135591  & 3.98 $10^{-7}$ & 1.81 $10^{-6}$  & $<$ 3.25 $10^{-6}$ \cr

&  5.50 $10^{-8}$  &  &  \cr

HD 150135   & & 1.43 $10^{-6}$  & 5.89 $10^{-6}$\cr

HD 150136   &   & 5.07 $10^{-6}$  & 4.01 $10^{-5}$ \cr

HD 163181   & 7.08 $10^{-6}$  & 6.32 $10^{-6}$  & 2.37 $10^{-6}$\cr

HDE 319718  &   & 2.90 $10^{-6}$  & $<$ 4.62 $10^{-6}$ \cr

&&&\cr
\hline
\multicolumn{3}{l} {1: computed from Vink et al. (2000)}
\end{tabular}
\end{center}
\end{table}

\subsection{Origin of the non-thermal emission}

Regarding the non-thermal sources detected here, possible origins
of the emission are accretion of matter onto compact companions,
shocks produced from line driven instabilities closer to the base
of the wind, and in massive binary systems, at the wind collision
region. Garmany et al. (1980) have estimated that at most 6\% of
all O stars could have an (optically) undetected companion.
Bieging et al. (1989) established that systems formed by an OB
star and a compact companion comprises less than 1\% of all
luminous OB stars. Being the frequency of compact companions so
low for early-type stars, the possibility of detecting emission
from this scenario becomes rather small.

HD 150136 is a known binary star and HD 124314, a possible
spectroscopic binary (see Table~1 for references). The
determination of the spectral classification of the secondary
stars by means of optical observations is crucial to decide
whether the non-thermal nature of these two candidates resides on
colliding winds.

CD--47$^{\circ}\,$4551 is not recognized either as a binary or
multiple  star but its radio image at 4.8 GHz exhibits a second
source. The J-band image from DSS (see Fig. 5) shows that a second
optical object could be present in the direction of this
candidate. There exists the possibility that the weak radio source
be originated by the presence of a companion of CD--47$^{\circ}$
4551, separated $\sim$ 9000 AU. A possible scenario could be the
following: the strongest non-thermal source, positionally
coincident with the star CD--47$^{\circ}\,$4551 could be generated
by inner shocks in the wind of the Of star or shocks at wind
collision region if the star has a close early-type companion. The
second (and also non-thermal) source could be emission from a
colliding winds region between the winds of CD--47$^{\circ}\,$4551
and a visual or wide-orbit object.

\begin{figure}
 \resizebox{\hsize}{!}{\includegraphics{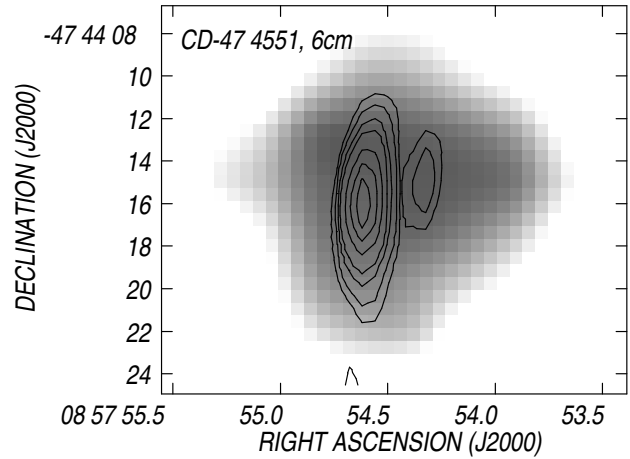}}
 \caption{Contour map of CD--47$^{\circ}\,$4551 at 6 cm superposed on
the J-band DSS image.}
  \label{fig5} \end{figure}

The proposal stated by Chapman et al. (1999) and Dougherty \&
Williams (2000) that non-thermal emission could generally indicate
wind interaction between a WR and a companion gives support to the
idea of the existence of a windy companion for the present
non-thermal sources, as Of stars are WR precursors. Optical
studies of CD--47$^{\circ}\,$4551 would be a fundamental tool to
investigate its binary status.

Unfortunately, the angular resolution of the current observations
does not suffice to discriminate if the non-thermal emission comes
from the wind of a star or from the site where two winds collide.
This kind of investigations must be carried out with very high
angular resolution telescopes, like MERLIN (see for example the
results of  Dougherty et al.
 2000 related to WR 146) and they are limited by the source declination.
In addition, variability studies would bring some information to
help addressing this question, in the sense that if changes in the
flux density are present, the existence of an early type
companion, and a wind-wind interaction region, is the most
plausible explanation for the observed non-thermal radio continuum
emission.

\section{Conclusions}

A number of conclusions can be drawn from the present
observations:

\begin{itemize}
\item  Five out of thirteen OB stars were detected at 3 and 6 cm
wavelengths. The three brightest stars, CD--47$^{\circ}$ 4551, HD
124314 and HD 150136, yielded negative spectral indices, giving
$<\alpha>\ \approx -0.8$, whereas HD 150135 and HD 163181 gave
positive spectral indices, $<\alpha>\ \approx +1.5$.

\item  None of the target stars, except HD 57060, had previously been
observed. This star presents now new (and lower) upper limits for
the flux density ($<$ 0.15 mJy at 3 cm) and for the mass loss rate
($< 3 \times 10^{-6}$ M$_\odot$ yr$^{-1}$). A previous value of
$\dot{M}$, derived from optical data was 8.5$\times 10^{-6}$
M$\odot$ yr$^{-1}$, and would correspond to a flux density of
about 0.6 mJy at 3 cm. Variations (a decrease in this particular
case) of the mass loss rate could explain the non detection above
0.15 mJy.

\item  More than half of the detected stars show spectral indices
consistent with contamination of non-thermal emission. If many
surveys of continuum emission are taken into account, the results
presented here increase the percentage of non-thermal stars up to
50\% among detected O-type stars. Non-thermal centimeter radio
emission may be a common phenomenon in Of stars. A similar
conclusion was reached by Chapman et al. (1999) concerning WR
stars. These authors state that the emission would come in most
cases from colliding winds.

\item Either shocks in the wind if the star has no early-type
companion, or shocks in the region where winds collide can explain
the non-thermal emission from three target sources. The stars
could have a massive companion: HD 150136 is a binary star, HD
124314 is a suspected binary, and CD--47$^{\circ}$ 4551 presents
both in radio and optical wavelengths a second source separated
5\arcsec.

\item  The flux density of HD 124314 was found to vary substantially.
Within two years the source almost duplicated its flux.

\item  In this small sample, no clear correlation between detectability
and distance, or spectral type, or luminosity class was found.
However, the detected stars were the most luminous, early-type and
closest candidates.

\item  The theoretical predictions recently published by Vink et al.
(2000) are considered a useful tool to estimate mass loss rates,
and can also help to foresee in a general way which stars are more
likely to be detected. Their predictions are in general agreement
with the results presented here for undetected and thermal
sources.

\end{itemize}

\begin{acknowledgements}
We are in debt to the referee, Dr. C. Leitherer, for valuable
comments. P.B. is grateful to the ATNF staff at Sydney and
Narrabri and wishes to thank Dr. G.E. Romero for useful
discussions, M. Johnston-Hollitt and R. Wark for assistance during
observations and FCAG and UNLP for travel support. We aknowledge
Dr. G. Bosch for his help with web images. This work has made use
of the SIMBAD database operated at CDS, Strasbourg, France, and
was partially supported by CONICET (under grant PIP 607/98).
\end{acknowledgements}

{}

\end{document}